\documentclass[journal]{IEEEtran}
\IEEEoverridecommandlockouts
\usepackage{cite}
\usepackage{amsmath,amssymb,amsfonts}
\usepackage{algorithmic}
\usepackage{graphicx}
\usepackage{textcomp}
\usepackage{xcolor}
\usepackage{subfigure}
\usepackage{multicol,multirow}
\usepackage{gensymb}

\def\BibTeX{{\rm B\kern-.05em{\sc i\kern-.025em b}\kern-.08em
    T\kern-.1667em\lower.7ex\hbox{E}\kern-.125emX}}
\begin{document}

\title{``One-Shot'' Reduction of Additive Artifacts in Medical Images}

\author{
Yu-Jen Chen$^{1}$, 
Yen-Jung Chang$^{1}$,
Shao-Cheng Wen$^{1}$,
Yiyu Shi$^{2}$,
Xiaowei Xu$^{3}$,\\
Tsung-Yi Ho$^{1}$,
Meiping Huang$^{3}$,
Haiyun Yuan$^{3}$, 
Jian Zhuang$^{3}$
\\
\textit{$^{1}$Department of Computer Science, National Tsing Hua University, Hsinchu, Taiwan\\ 
$^{2}$Department of Computer Science and Engineering, University of Notre Dame, IN, USA\\
$^{3}$Guangdong Provincial People's Hospital, Guangdong Acadamic of Medical Science, Guangzhou, China
}

\thanks{This research was approved by the Research Ethics Committee of Guangdong General Hospital, Guangdong Academy of Medical Science with the protocol No. 20140316.}
}

\maketitle
\begin{abstract}
Medical images may contain various types of artifacts with different patterns and mixtures, which depend on many factors such as scan setting, machine condition, patients’ characteristics, surrounding environment, etc. However, existing deep-learning-based artifact reduction methods are restricted by their training set with specific predetermined artifact types and patterns. As such, they have limited clinical adoption. In this paper, we introduce One-Shot medical image Artifact Reduction (OSAR), which exploits the power of deep learning but without using pre-trained general networks. Specifically, we train a light-weight image-specific artifact reduction network using data synthesized from the input image at test-time. Without requiring any prior large training data set, OSAR can work with almost any medical images that contain varying additive artifacts which are not in any existing data sets. In addition, Computed Tomography (CT) and Magnetic Resonance Imaging (MRI) are used as vehicles and show that the proposed method can reduce artifacts better than state-of-the-art both qualitatively and quantitatively using shorter test time. 
\end{abstract}

\section{Introduction}
\label{sec:introduction}
Deep learning has demonstrated its great power in artifact reduction, a fundamental task in medical image analysis to produce artifact-free images for clinical diagnosis, decision making, and accurate quantitative image analysis. Most existing deep-learning-based approaches
use training data sets that contain paired images (identical images with and without artifacts) to learn the distribution of additive artifact features, such as Gaussian noise, Poisson noise, motion artifact, etc. e.g. Yang et al. \cite{yang2018low} adopted Wasserstein distance and perceptual loss to ensure the similarity between input and the generated image (WGAN-VGG), and Kang et al. \cite{kang2018cycle} used cycle-consistent adversarial denoising network (CCADN) that learns the mapping between the low- and routine-dose cardiac phase without matched image pairs. As artifact-free images are usually hard to obtain clinically, simulations are often involved in establishing such data sets, i.e., superposing the predefined additive noise to images. However, the simulated noise patterns may be different from those in real situations, thus leading to biased learning \cite{yang2018low,veraart2016denoising}. To deal with this issue, Chen et al. \cite{chen2020zero} proposed a Zero-Shot medical image artifact reduction (ZSAR) approach, which utilized an unsupervised clustering method to extract the artifact pattern and restore the original images. However, their improvement are limited in the scenarios where the intensity difference between the artifact and the edge of the tissue is not large. In such cases, the clustering algorithm could not recognize the artifact pattern well.

To address these issues, we propose a ``One-Shot'' image-specific artifact reduction framework (OSAR) for additive noise or artifact, which exploits the power of deep learning model yet does not require any clean image reference or a large pre-defined training data set. By focusing on those additive artifacts which are laid above a uniform background, it is almost always possible to extract artifact patterns from the given image. Our method simply takes an image with artifacts as input and applies both training and test procedure to the input image for artifact reduction. The proposed framework requires only a few radiologist-annotated regions-of-interest (ROIs) in the image to train a small-scale Convolutional Neural Network (CNN). This CNN is then used to automatically recognize the area with artifacts to synthesize paired patches from the input image. Afterward, these paired data are used to train another light-weight network that reduces the artifacts in the image. To tackle the issues of the limited synthesized data size, and the requirement of fast test-time training, we designed a compact attentive-artifact-reduction-network that pays special visual attention to the regions with artifacts and restores obstructed information. 

Experimental results using clinical Computed Tomography (CT) and Magnetic Resonance Imaging (MRI) data show that the proposed approach, OSAR, outperforms the state-of-the-art in both qualitatively and quantitatively by a large margin when test images are affected by extra artifacts that are not in the training set. Even when test on images that have similar artifacts as the training data, OSAR can still work better.

The name ``One-Shot'' is borrowed from the classical image classification problem where only a single image is used for training. Our work here also trains on only one specific input image, though we test on the same image as input, we could still consider as ``One-Shot'' learning. Also, although our work requires annotating ROIs in each input image by radiologists, our ablation study shows that only a very small number of ROIs are sufficient.


Even though manufacturers could include artifact-reducing algorithm to improve image quality for machines, scan setting, machine condition, patients' characteristics, surrounding environment can all affect the image quality. Therefore, there are a large body of existing work on medical image artifact reduction after the images are captured, such as \cite{chen2020zero, kang2018cycle, wolterink2017generative, yang2018low}, and the motivation of our work follows these work.

The main contributions of the proposed method are as follows: 
\begin{itemize}
    \item It can handle various types of additive artifacts with different patterns and their mixtures on a uniform background.
    \item It does not require any pre-training on large pre-defined training data set and can run with modest amounts of computation resources. 
    \item It outperforms state-of-the-art both qualitatively and quantitatively in both CT and MRI.
\end{itemize}

\section{Related Works}
\label{sec:related_works}
In this paper, we limit our discussion to CT and MRI as they are the vehicles to demonstrate the effectiveness of our method in the experiments. For CT, artifacts can be classified into patient-based (e.g., motion artifact), physics-based (e.g., Poisson noise), and helical and multi-channel artifacts (e.g., cone beam effect) according to the underlying cause \cite{boas2012ct}. For MRI, such noise types as truncation artifacts, motion artifacts, aliasing artifacts, Gibbs ringing artifacts, etc. \cite{krupa2015artifacts}, are common in real-world scenario. These artifacts are caused by a number of factors, including scan setting, machine condition, patient size and age, surrounding environment, etc. These artifacts may occur at random places in an image. In addition, multiple artifacts can occur and mix in the same image. Although some general-purpose denoising methods such as Deep Image Prior \cite{ulyanov2018deep} and non-deep-learning-based methods such as BM3D \cite{dabov2006image} and NLM \cite{buades2005non} can also be readily applied, their results are inferior in this specific problem. Thus, we limit our discussion to deep-learning-based methods on medical image only.

For noise artifacts on CT images, Chen et al. \cite{chen2017low} proposed a Convolution Neural Network (CNN) to reduce the Poisson noise on low-dose CT images and reconstruct the corresponding routine-dose CT images. Wolterink et al. \cite{wolterink2017generative} designed a Generative Adversarial Network (GAN) with CNN for low-dose CT images Gaussian noise reduction. Yang et al. \cite{yang2018low} adopted Wasserstein distance and perceptual loss to ensure the similarity between input and the generated image. As for the MRI images, Manj´on and Coupe \cite{manjon2018mri} proposed a simple CNN network for 2D MRI artifact reduction and Jiang et al. \cite{jiang2018denoising} explored multi-channel CNN for 3D MRI Rician noise denoising. However, most of the existing approaches still require simulations to generate the paired data, which may lead to biased learning when simulated artifacts are different from real ones. To eliminate the need for paired training data, recently Noise2Noise-based \cite{lehtinen2018noise2noise} methods have been developed, where the denoising networks are learned by mapping a noisy image to another noisy realization of the same image. Kang et al. \cite{kang2018cycle} used cycle-consistent adversarial denoising network (CCADN) that learns the mapping between the low- and routine-dose cardiac phase without matched image pairs. Wu et al. \cite{wu2019consensus} proposed a consensus neural network to enhance the performance of Noise2Noise and applied it on medical images. 

However, all these methods are constrained by their specific training data, which can hardly capture all possible artifact types and patterns that since they may vary and mix. As such, all these trained frameworks may only have limited clinical use.
To deal with similar issue, Chen et al. \cite{chen2020zero} proposed a Zero-Shot medical image artifact reduction (ZSAR) approach, which utilized the an unsupervised clustering method, K-means, to extract the artifact pattern and restore the original images. However, their improvement are limited in the scenarios where the intensity difference between the artifact and the edge of the tissue is not large. In such cases, the clustering algorithm could not recognize the artifact pattern well.

\section{Methods}
\label{sec:methods}
\begin{figure*}[t]
\begin{center}
\includegraphics[width=0.85\linewidth]{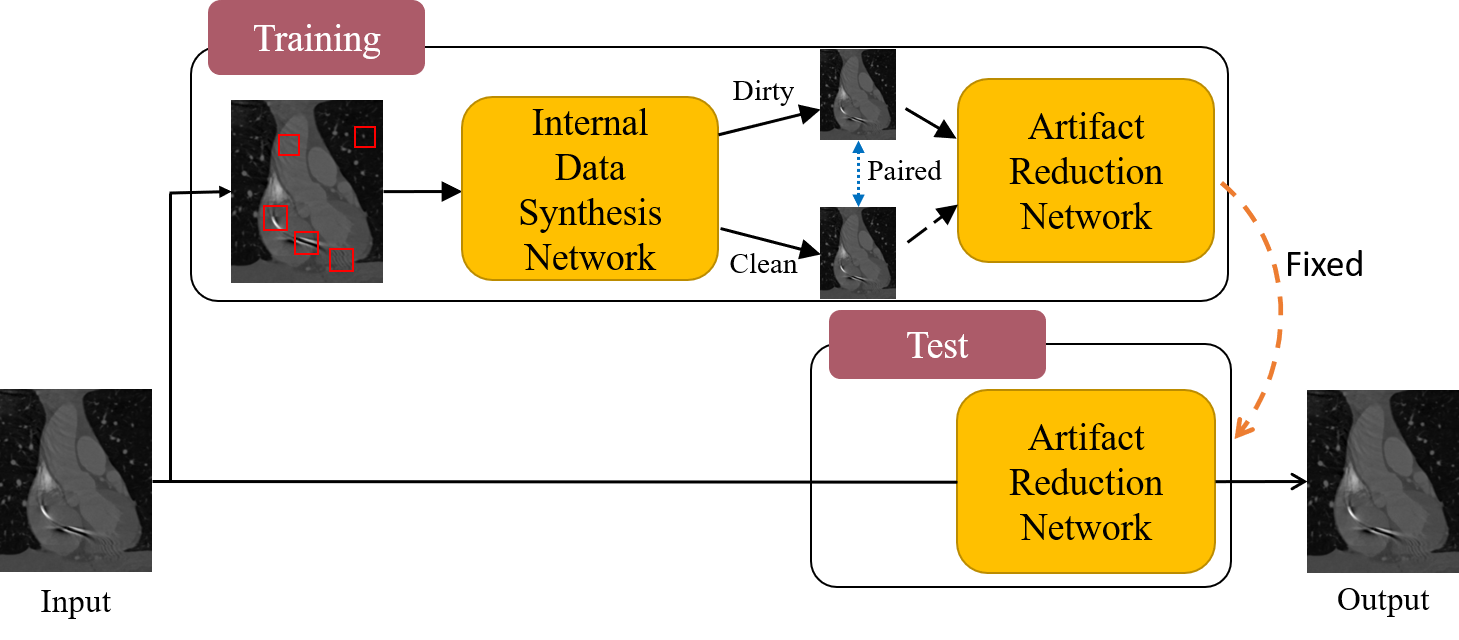}

\caption{Overall structure of the proposed OSAR, which mainly contains two parts: internal data synthesis network (IDSN) and attentive artifact reduction network (AARN). The trained AARN network is directly used in test, though the input image size is different (patches with sized 32$\times$32 for training v.s. full image size for test). 
}
\label{Fig.flow}
\end{center}
\end{figure*}

Fundamental to our approach is the fact that we can find an area with additive artifacts on a relatively uniform background in most medical images. 
This provides the possibility to synthesize paired noise-affected and noise-free training patches from an image with artifacts. Thus, the deep neural network could recognize the artifact distribution from the synthesized data set. In addition, since both the training set and test set come from the same image, the associated artifact reduction network can be compacted, and there is no overfitting concern. 

For clarity, we call the phase where the networks are trained to optimize their weights based on the input image as ``training'', and the phase where the trained network is inferred to that image to reduce artifacts as ``test''. We would like to emphasize that both training and test are done on the spot for the specific input. 
The overall architecture of the proposed OSAR framework is shown in Fig. \ref{Fig.flow}. It takes in a 2D image and uses a limited number of ROIs annotated by radiologists to train an Internal-Data-Synthesis-Network (IDSN) for artifact pattern extraction. 
The paired data generator then synthesizes a large number of paired patches from the extracted artifact patterns, and is further used to train an Attentive-Artifact-Reduction-Network (AARN).

\begin{figure}[t] \centering
\includegraphics[width=0.97\linewidth]{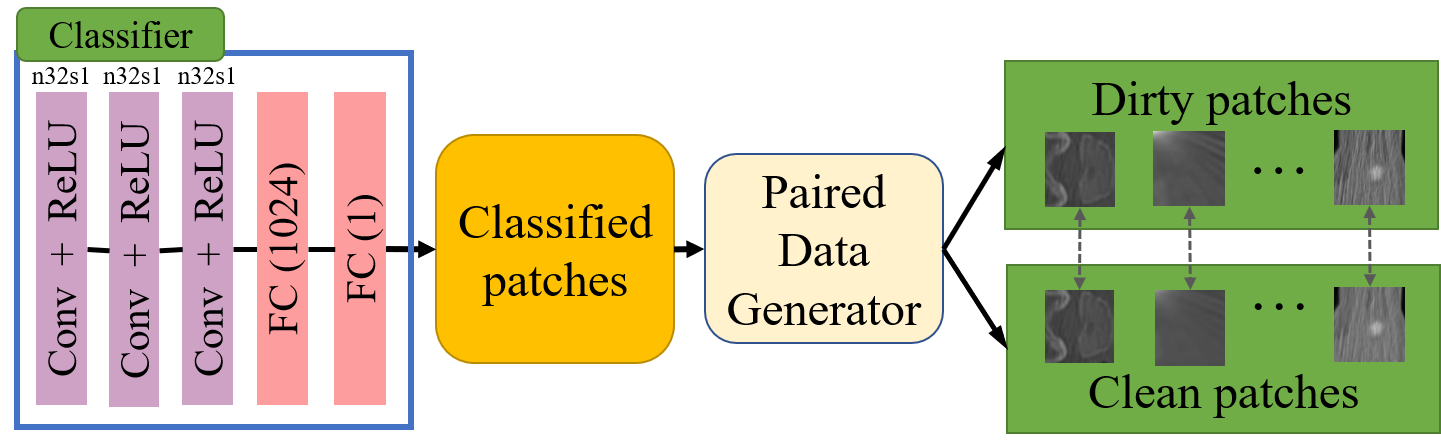}
\caption{The structure of internal data synthesis network (IDSN) contains a light-weight classifier with only three convolution layers and two fully-connected layers and a paired data generator algorithm.}
\label{Fig.IDSN}
\end{figure}

\subsection{Internal-Data-Synthesis-Network (IDSN)}



A graphical illustration of the IDSN is shown in Fig. \ref{Fig.IDSN}. 
The proposed IDSN contains a light-weight CNN-based classifier that recognizes the patches as either artifact or the other, and a paired data generator then synthesizes the paired data for further use.
Note that different medical images have different ranges of pixel values, we normalize the pixel value of each slice to [0,1] before processing and scale them back afterwards.  

\begin{figure}[h]
\begin{center}
\subfigure[]{\label{Fig.A-type_1}
\includegraphics[width=0.2\linewidth]{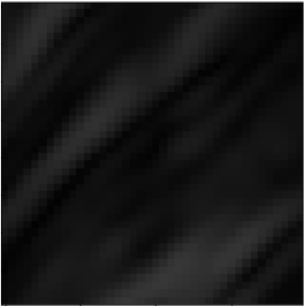}}
\subfigure[]{\label{Fig.A-type_2}
\includegraphics[width=0.2\linewidth]{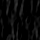}}
\subfigure[]{\label{Fig.A-type_3}
\includegraphics[width=0.2\linewidth]{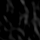}}
\subfigure[]{\label{Fig.A-type_4}
\includegraphics[width=0.2\linewidth]{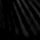}}

\caption{Examples for A-type patches recognized by IDSN. A-type patches contain artifacts on a uniform background such as tissues, air, etc, which is considered as artifact patches.}
\label{Fig.Categories}
\end{center}
\end{figure}

Specifically, the classifier has five layers and is designed to rapidly classify a patch into two categories: 1). \textbf{A-type}, which contains artifacts on a uniform background such as tissues, air, fat, etc (artifact patch) and 2). \textbf{N-type} for all the remaining ones (non-uniform, with or without artifacts). 
Examples for the A-patches are shown in Fig. \ref{Fig.Categories}.
To train the CNN, radiologists manually pre-annotated a few ROIs of each type (with size 32$\times$32) in the input image. These ROIs are then augmented to form the training data. The standard softmax cross-entropy loss is used for training. After the model is trained, we slice the input image into 32$\times$32 patches and apply the trained CNN to classify each of them into one of the two types. Our experiments suggested that only a few ROIs are needed since the following artifact reduction network is tolerant to classification errors introduced in IDSN: in all the images we tested, 7 annotated ROIs are sufficient to yield a classification accuracy round 80\%, while increasing it to 27 only slightly boosts the accuracy. 

A paired data generator then extracts the artifact patterns from A-type patches by subtracting the mean pixel value of each patch. Next, it synthesizes paired data by superposing these patterns to all the patches. Each superposition will generate a pair of patches (``clean'' and ``dirty''). 
We refer to the one before superposition as ``clean'' patch, while the one after as ``dirty'' patch.
Note that the clean patch may still contain pre-existing artifacts (e.g., the A-type ones). However, we find that as long as the corresponding dirty patch has higher artifact density, such a dirty-clean pair is still effective in training the AARN. Similar concept is proved in Noise2Noise-based \cite{lehtinen2018noise2noise} approaches. We also randomly select some of the A- and N-type patches and use them to form identical dirty-clean pairs (same patch as both dirty and clean) to reflect the fact that not all areas in an image contain artifacts.

Unlike previous simulation-based approaches, the synthesized dirty patches have artifacts that completely resemble the artifacts in the exact image, thus eliminating any biases. 

\subsection{Attentive-Artifact-Reduction-Network (AARN)}
\label{sec:aarn}

\begin{figure*}[ht]
\begin{center}
\includegraphics[width=0.95\linewidth]{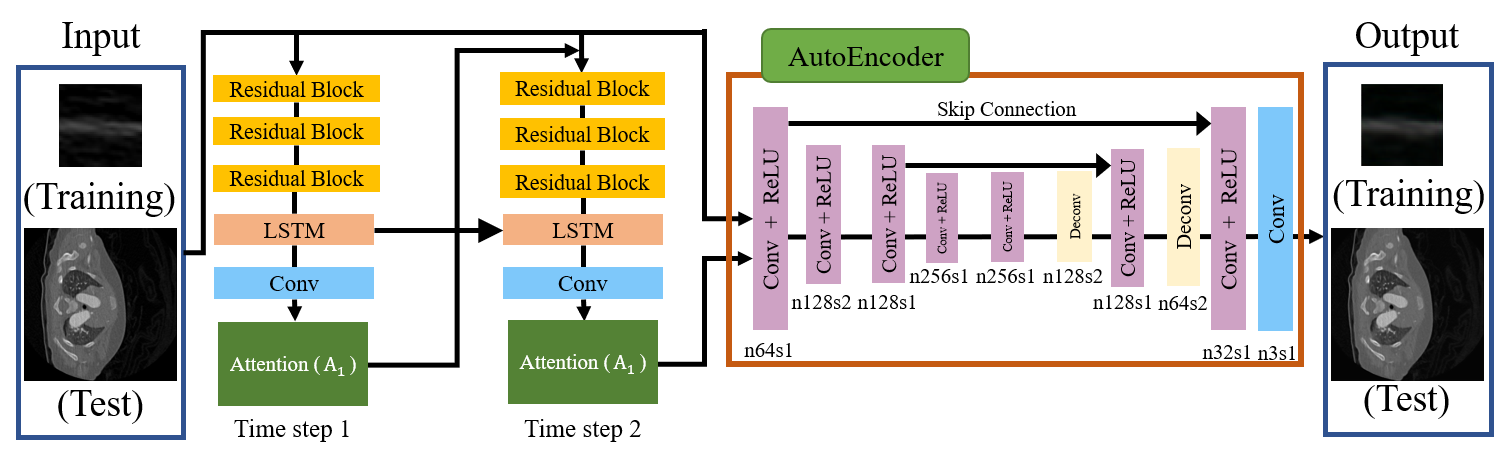}
\caption{The architecture of AARN, composed of an attentive-recurrent network with two time steps followed by an autoencoder. The attention maps $A_1$ and $A_2$ are used to help the autoencoder focus on the regions with artifacts. The input is a patch with size 32$\times$32 during training and the full input image during test.}
\label{Fig.generator}
\end{center}
\end{figure*}


After synthesizing the paired patches, theoretically, any existing supervised CNN-based artifact reduction networks can be trained. However, a key issue here is that we perform the solution on each input image. A deep and complex network may need a large number of paired data and take a long training time. On the other hand, small networks may not attain desired performance. 

Attentive-generative-network was first introduced in \cite{qian2018attentive} for raindrop removal, which injects visual attention to dirty areas for faster and more accurate information restore.
In this work, we significantly simplified the network structure and removed the part of the adversarial network to allow test-time training with few training data.


As shown in Fig. \ref{Fig.generator}, the AARN is formed by a two-step attentive-recurrent-network followed by a 10-layer contextual autoencoder to reduce artifacts and to restore the information obstructed by them. Each block in the recurrent network extracts features from its input and feeds the generated feature map (attention map) to the next block. We also create a binary map $M$ by calculating the pixel-wise difference between pairs of dirty and clean patches from IDSN.
A artifact threshold is set to 0.01, which $M[x] = 1$ if the distance in the corresponding values of pixel $x$ is greater than the threshold, indicating $x$ is in a dirty region. Otherwise, $M[x] = 0$ if the pixel $x$ is in the clean region.  

\begin{figure}[hbtp]
\begin{center} 
\subfigure[Input]{\label{Fig.6Input}
\includegraphics[width=0.3\linewidth]{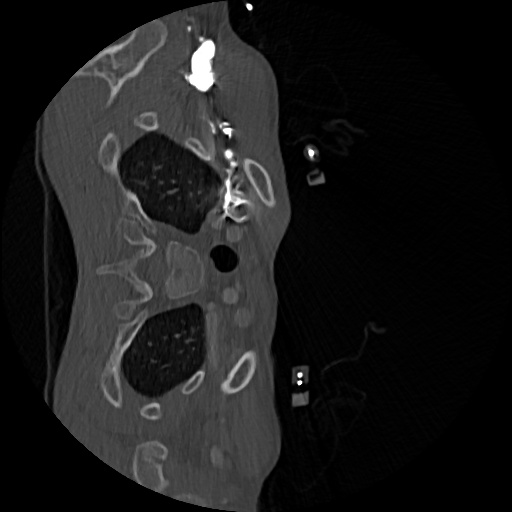}}
\subfigure[Time step 1]{\label{Fig.mask1}
\includegraphics[width=0.3\linewidth]{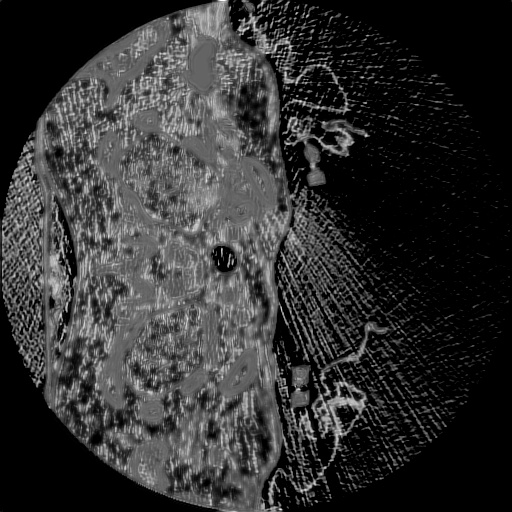}}
\subfigure[Time step 2]{\label{Fig.mask2}
\includegraphics[width=0.3\linewidth]{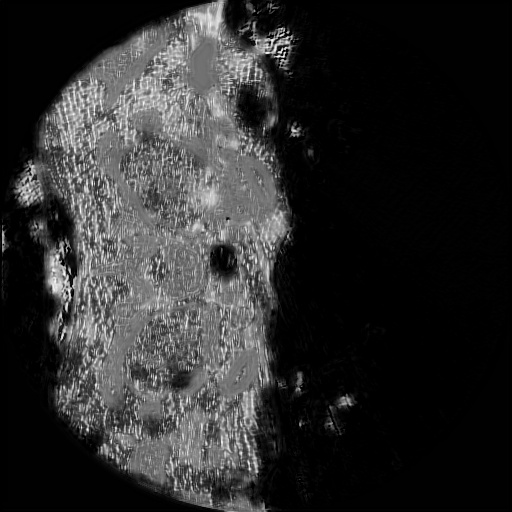}}

\caption{Input image and the attention maps generated by the attentive-recurrent-network during test.}
\label{Fig.Mask}
\end{center}
\end{figure}

We would like the attention map to be as close to the binary map as possible. As such, the loss function $L_{ATT}$ in each recurrent block calculates the mean square error (MSE) between the attention map $A_t$ at the two time steps ($t=1$ and $t=2$) and the binary map $M$ as
\begin{small}
\begin{equation}
    L_{ATT}(A_t,M)= 0.8 \cdot L_{MSE}(A_1,M)\ +  L_{MSE}(A_2,M).
\end{equation}
\end{small}
Examples of the attention map can be found in the Fig. \ref{Fig.Mask}. 

After the attention map is generated, it is combined with the input of the recurrent network to form the input of the contextual autoencoder, which then generates an image with artifacts reduced.
The loss function related to the autoencoder is the multi-scale loss.

The multi-scale loss extracts features from different decoder layers which help capture more contextual information from different scales:
\begin{small}
\begin{equation}
    L_M(\{F\},\{T\})=\sum_{i} w_i \cdot L_{MSE}(F_i,T_i)
\end{equation}
\end{small}
where $F_i$ is the feature extracted from the $i$-th autoencoder layers, $w_i$ is the weight, and $T_i$ is the ground truth from the corresponding clean patch at the same scale. 
Through experiments, we find that using outputs of the decoder in each resolution (layer 5, 7, and 9 of the autoencoder) yields the best results. Note that the later layer has a larger weight $w_i$, and the last layer has the weight set as 1.0 (0.6, 0.8, and 1.0, respectively).

Eventually, the finally loss function for the AARN model $L$ can be fomulated as: 
\begin{small}
\begin{equation}
    L =L_{ATT}+L_M\\ \label{eq:7}
\end{equation}
\end{small}
Throughout our experiments, we find that these two loss term are balanced and work excellently for additive artifact on both CT and MRI images.  Moreover, the training of light-weight AARN takes at most four epochs to converge, which is much faster than ZSAR \cite{chen2020zero}, CCADN \cite{kang2018cycle}, and BM3D \cite{dabov2006image} (Please see Section \ref{sec:results} for more detail runtime comparison). This advantage is critical for test-time training.



\section{Experiments}
\label{sec:experiments}

\begin{figure*}[t]
\centering
\includegraphics[width=0.98\linewidth]{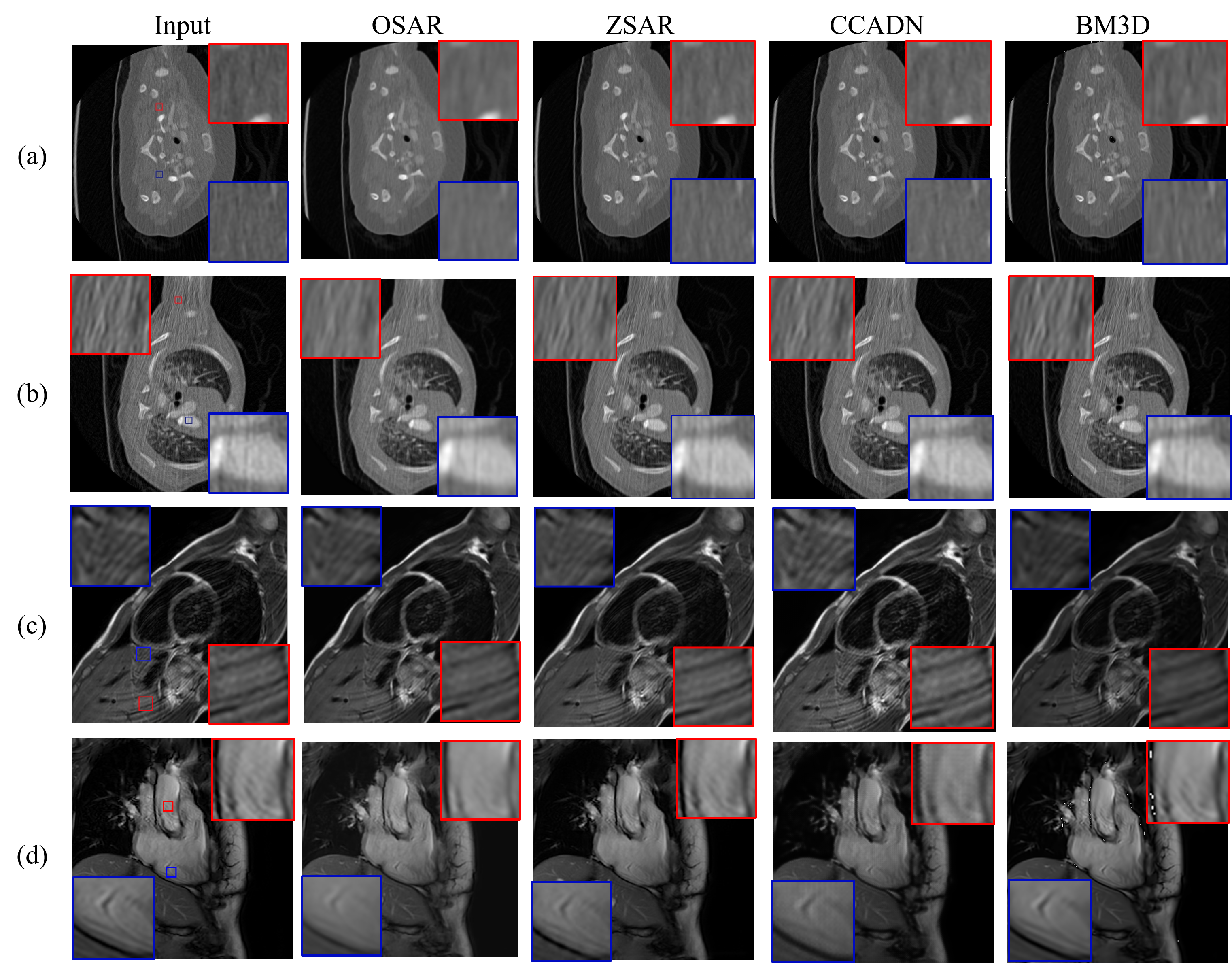}
\caption{Qualitative comparison for additive artifact reduction in cardiac CT images and MRI images by various methods. Artifact patterns for ideal scenario cases (a) and (c) appeared in the training set, and that non-ideal scenario cases (b) and (d) did not. 
Quantitative comparisons for the largest homogeneous areas inside the marked regions can be found in Table \ref{Table:Exp}.}
\label{Fig.Exp}
\end{figure*}

\subsection{Cardiac Data Set and Evaluation Metrics}
Our dataset includes 24 series of CT images, and 16 series MRI data.
For CT, all examinations were performed by our wide detector 256-slice MDCT scanner with 8 cm of coverage, using the following protocol: collimation, (96-128)$\times$0.625 mm; rotation time, 270 ms, which corresponds to a 135-ms standard temporal resolution; slice thickness, 0.9 mm; reconstruction interval, 0.45 mm. Adaptive axial z-collimation was used to optimize the cranio-caudal length. Data were obtained at 40-50\% of the RR interval, utilizing a 5\% phase tolerance around the 45\% phase.The dosages are between 80 kVp/55 mAs and 100 kVp/651 mAs, with such low dose circumstances we can capture the Poisson noise from the CT images.

MRI was performed on our 3T system. 
Along with long-axis planes, a stack of short-axis single-shot balanced standard steady-state in free-precession sequence images from apex to basal was collected. The imaging parameters were as follows: field of view, 230 mm$\times$230 mm; voxels, 2 mm$\times$2 mm$\times$8 mm; repetition time, (3.0-3.2) ms; echo time, (1.5-1.6) ms; sense factor, 2; minimum inversion time, 105 ms; and flip angle, 45\degree.
The motion artifact is captured in the dataset.

All CT and MRI images were qualitatively evaluated by our radiologists on structure preservation and artifact level. For quantitative evaluation, due to the lack of ground truth, for both CT and MRI we followed most existing works \cite{wolterink2017generative,yang2018low,goerner2011measuring,kellman2005image} and selected the most homogeneous area in regions of interest selected by radiologists. We divided the mean of the pixel values in the most homogeneous areas by their standard deviation and used the resulting Signal-to-Noise ratio (SNR) as the metric. Remind that the mean (substance information) discrepancy after artifact reduction should not be too large.

\subsection{Methods and Training Details}
For OSAR, we applied the Adam optimization \cite{kingma2015adam} method to train both IDSN and AARN by setting the learning rate to 0.0005. 

The paired data generator produced 100,000 paired patches. The maximum number of epochs was set to 4 and the batch size is set to 270 for AARN training, but in most cases, it converged within only two or three epochs. 
Xavier initialization \cite{glorot2010understanding} was used for all the kernels.
Only one patient's data (a single image) was used for each training and test. For each image, our radiologists annotated around 7 ROIs (the impact of the number of the ROIs will be discussed in Section \ref{sec:ablation}). 

We compared OSAR with the state-of-the-art deep-learning-based medical image artifact reduction methods CCADN \cite{kang2018cycle} and ZSAR \cite{chen2020zero}, and trained each following exactly the same settings reported.
The CT and MRI training data sets for CCADN contain 100,000 image patches generated from a large number of patients scanned by the systems described above, using simulation when necessary, to ensure sufficient variability and representativeness. 

We also compared OSAR with another state-of-the-art general-purpose non-learning-based algorithms, BM3D. For each image, we tuned the parameters in these methods such as template window size and searching window size to attain the best quality.

\section{Results}
\label{sec:results}

\begin{table*}[h]
\caption{Mean and Signal-to-Noise Ratio (SNR) for the largest homogeneous areas inside the marked regions of the CT images in Fig. \ref{Fig.Exp} (in Hounsfield Unit) (a) and (b). Mean and SNR for the largest homogeneous areas inside the marked regions of the MRI images in Fig. \ref{Fig.Exp} (c) and (d).}
\label{Table:Exp}
\centering
\begin{tabular}{c|c|ccccccc}
\hline
Modality & Scenario & Case & & Input & OSAR & ZSAR & CCADN & BM3D\\
\hline
\multirow{8}{*}{CT} & \multirow{4}{*}{Ideal} & \multirow{2}{*}{(a)Red} & Mean & 54.0 & 94.5 & 61.9 & 68.2 & 48.8\\ 
                         &&& SNR & 0.68 & \textbf{2.35} & 0.98 & 1.03 & 0.67\\
\cline{3-9}
&&\multirow{2}{*}{(a)Blue} & Mean & 104.4 & 146.9 & 98.2 & 118.5 & 99.3\\ 
                         &&& SNR & 1.58 & \textbf{4.85} & 1.91 & 2.42 & 1.86\\
\cline{2-9}
\cline{2-9}
& \multirow{4}{*}{Non-Ideal} &\multirow{2}{*}{(b)Red} & Mean & 82.9 & 121.9 & 95.8 & 96.8 & 78.5\\ 
                         &&& SNR & 0.57 & \textbf{1.20} & 0.85 & 0.75 & 0.55\\
\cline{3-9}
&&\multirow{2}{*}{(b)Blue} & Mean & 790.4 & 851.6 & 827.9 & 823.9 & 786.0\\ 
                         &&& SNR & 8.84 & \textbf{11.89} & 11.20 & 10.60 & 9.36\\
\hline
\hline
\multirow{8}{*}{MRI} & \multirow{4}{*}{Ideal} & \multirow{2}{*}{(c)Red}  & Mean & 477.3 & 479.7 & 494.9 & 547.0 & 476.1\\ 
                         &&& SNR  & 6.2 & \textbf{7.5} & 7.1 & 7.0 & 6.2\\ 
\cline{3-9}
&&\multirow{2}{*}{(c)Blue} & Mean & 381.9 & 396.0 & 396.1 & 449.0 & 380.3\\
                         &&& SNR & 7.6 & 10.8 & \textbf{11.7} & 10.5 & 7.8\\
\cline{2-9}
\cline{2-9}
& \multirow{4}{*}{Non-Ideal} &\multirow{2}{*}{(d)Red}  & Mean & 842.6 & 836.7 & 830.0 & 779.5 & 839.8\\ 
                         &&& SNR & 17.4 & \textbf{33.8} & 21.1 & 19.0 & 19.5\\
\cline{3-9}
&&\multirow{2}{*}{(d)Blue} & Mean & 633.7 & 646.2 & 636.2 & 592.9 & 631.0\\
                         &&& SNR & 8.9 & \textbf{15.5} & 13.0 & 10.4 & 9.4\\
\hline
\hline
\multicolumn{2}{c|}{Ideal} & \multirow{2}{*}{Overall}  & $\Delta$SNR & 0.0\% & \textbf{128.9\%} & 33.3\% & 38.9\% & 5.4\% \\ 
\multicolumn{2}{c|}{Non-Ideal} &  & $\Delta$SNR & 0.0\% & \textbf{78.3\%} & 35.7\% & 19.3\% & 6.7\% \\
\hline
\multicolumn{3}{c}{Overall}  & $\Delta$SNR & 0.0\% & \textbf{103.6\%} & 34.5\% & 29.1\% & 6.0\% \\
\hline
\end{tabular}
\end{table*}

In this section, we report the comparisons between OSAR, ZSAR, CCADN, and BM3D in the ideal scenario and the non-ideal scenario and compare the average execution times among all four methods. In the ideal scenario, the artifact pattern in the test set also appeared in the training set, while the non-ideal scenario did not.

\subsection{Experimental Results Comparison with State-of-the-art in Ideal Scenario}
We start our discussion with the ideal scenario where the artifact in both training set and test set contain Poisson noise only. The qualitative results for ZSAR, CCADN, BM3D, and OSAR are shown in Fig. \ref{Fig.Exp}(a). All the four methods preserved structures well and OSAR had smaller noise than the other three visually. This is expected as OSAR was trained on the specific image and thus more effective in reducing the noise contained therein. Our radiologists then selected the largest homogeneous areas inside the regions marked with red and blue rectangles for quantitative comparison, and the results are summarized in Table \ref{Table:Exp} (case a). From the table, OSAR achieved the highest SNR which is about 2 to 3$\times$ larger then the other three method.

We further applied the four methods to MRI motion artifact reduction in the ideal scenario that the test MRI image only contains motion artifact pattern similar to that in the training set. The qualitative results are shown in Fig. \ref{Fig.Exp}(c). Though all the methods preserved structures well, OSAR led to the best motion artifact reduction. The corresponding statistics for the largest homogeneous areas inside the marked regions are in Table \ref{Table:Exp} (case c). Although CCADN achieved almost the same SNR as OSAR, it had large mean discrepancy, which was about 14\%. As for BM3D, it preserve the best on mean information. However, the SNR performance is almost the same as input, which the improvement is minor.

\subsection{Experimental Results Comparison with State-of-the-art in Non-Ideal Scenario}
Next, we studied the non-ideal scenario where different artifact patterns or strength of artifacts absent from the training set appeared in the test image. For CT denoising, the qualitative results are shown in Fig. \ref{Fig.Exp}(b) and the corresponding mean and SNR numbers are presented in Table \ref{Table:Exp} (cases b). As for the results for MRI with different motion artifact patterns are shown in Fig. \ref{Fig.Exp}(d) and Table \ref{Table:Exp} (case d), respectively.

Qualitatively, we can see that in both CT and MRI images, the stripe shown up in the red and blue regions of OSAR is much smoothing than ZSAR, CCADN and BM3D, which the artifacts (stripe) are reduced properly. In addition, in Fig. \ref{Fig.Exp}(d), BM3D obtain several white spots in the red regions, which is unacceptable. Quantitatively, for CT images, OSAR outperformed ZSAR, CCADN and BM3D, achieving up to 41\%, 60\% and 118\% higher SNR, respectively. 
For MRI motion artifact reduction, all four methods kept the mean value well. However, OSAR attained up to 60\%, 77\%, and 73\% higher SNR than ZSAR, CCADN, and BM3D, respectively.

\subsection{Overall Results Comparison with State-of-the-art}
In the Table \ref{Table:Exp}, we also report the overall statistic results for all the test cases in CT and MRI images for ideal and non-ideal scenarios. The $\Delta$SNR represent the SNR improvement compared with the input image, which should be maximized. 

From the table, we can see that the OSAR has the optimal overall SNR improvement in ideal scenarios. Moreover, the SNR improvement for OSAR is about 3$\times$ larger than ZSAR and CCADN, and 20$\times$ larger than BM3D. 
As for the non-ideal scenario, we obtained a similar improvement trend with the ideal scenario. The SNR improvement for OSAR is about 2$\times$ larger than ZSAR, 4$\times$ larger than CCADN, and 10$\times$ larger than BM3D.

To summarize the results, OSAR obtain the optimal SNR improvement, which is 103.6\% in overall cases; that is, the proposed method reduces additive artifacts in both ideal and non-ideal scenarios better than the state-of-the-art.

\subsection{Execution Time Comparison}

\begin{figure}[h]
\begin{center}
\includegraphics[width=0.9\linewidth]{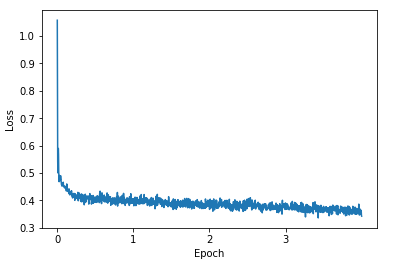}

\caption{The plot of the training loss for Fig. \ref{Fig.Exp}(b). The loss shows that our model usually converges within 2-3 epochs.}
\label{Fig.Epoch_impact}
\end{center}
\end{figure}

To show that test time training is feasible, as shown in Table \ref{Table:Time}, we compared the average execution times of OSAR with ZSAR, CCADN (which only include test), and BM3D on the CT and MRI images above. 
From the table, OSAR needs shorter runtime than ZSAR, CCADN, and BM3D. Since ZSAR used an iterative method for artifact reduction, it takes more time on the test phase. The speed of OSAR is brought by two facts: 1) In Fig. \ref{Fig.Epoch_impact}, we can see that the training loss usually converges within 2-3 epochs, where more epochs would not have a significant improvement.  2) It is much simpler than CCADN in structure and thus takes less time to process each 2D image of the 3D series.

\begin{table}[h]
\caption{Average execution time comparison for a 3D series of CT (512$\times$512) and MRI images (320$\times$320) (in seconds).}
\begin{center}
\label{Table:Time}
\begin{tabular}{|c|c|c|c|c|}
\hline
& OSAR & ZSAR & CCADN & BM3D\\
\cline{2-5}
 & Train+Test & Train+Test & Test & Test \\
\hline
CT (484 slices) & 404+726 & 360+2758 & 3533 & 1868\\
\hline
MRI (360 slices) & 401+468 & 150+1021 & 1294 & 1188\\
\hline
\end{tabular}
\end{center}

\end{table}

\begin{figure}[h]
\begin{center}
\setlength{\tabcolsep}{2pt}
\begin{tabular}{cccc}
\includegraphics[width=0.23\linewidth]{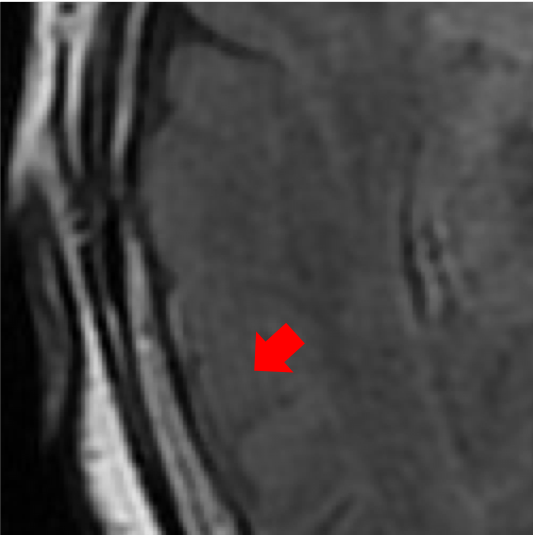} &
\includegraphics[width=0.23\linewidth]{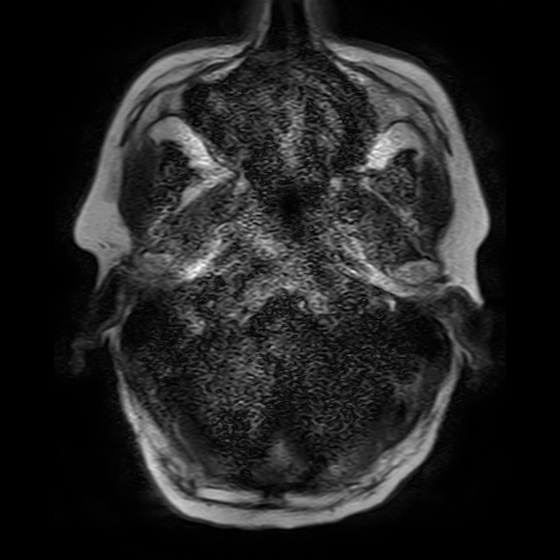} &
\includegraphics[width=0.23\linewidth]{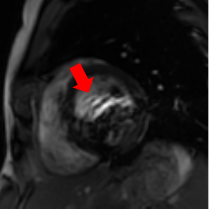} &
\includegraphics[width=0.23\linewidth]{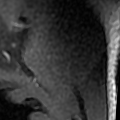} \\
\includegraphics[width=0.23\linewidth]{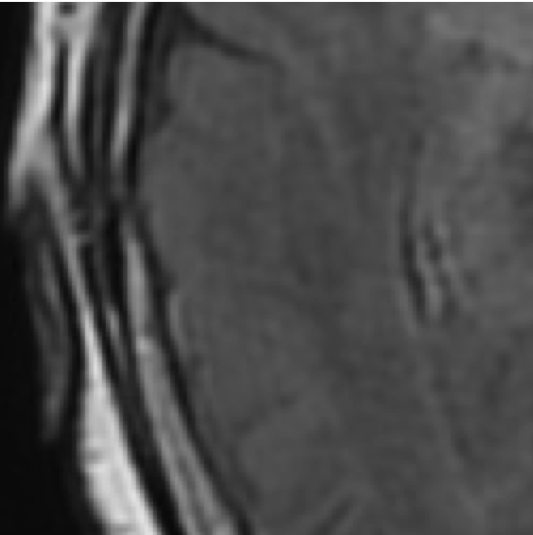} &
\includegraphics[width=0.23\linewidth]{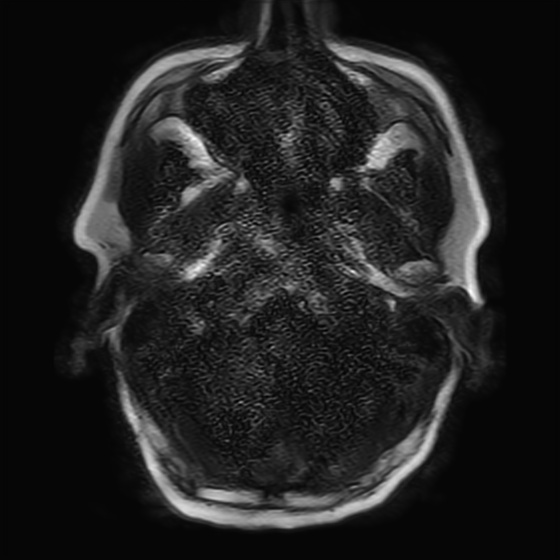} &
\includegraphics[width=0.23\linewidth]{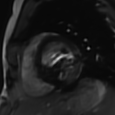} &
\includegraphics[width=0.23\linewidth]{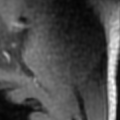} \\
(a) & (b) & (c) & (d) \\
\end{tabular}
\caption{OSAR applied to other types of artifacts on MRI image. (a) Gibbs ringing artifacts, (b) aliasing artifact, (c) spatially variant noise, and (d) intensity non-uniformity.}
\label{Fig.MRI_extra_artifact}
\end{center}
\end{figure}

\subsection{Results on Other Types of MRI Artifacts} \label{sec:other_types}
In this section, we show how OSAR performs on other types of additive artifacts for MRI, including Gibbs ringing artifacts, aliasing artifact, spatially variant noise, and intensity non-uniformity. Due to different environment circumstances, our training dataset cannot contain all types of artifact for model recognition. However, the proposed method successfully deal with the issue, which fit the clinical use. From the Fig. \ref{Fig.MRI_extra_artifact}, it is clear that OSAR is still effective to handle them, even though a training data set containing these artifacts is not available.

\section{Ablation Study}
\label{sec:ablation}
In this section, we conduct the ablation studies on the effectiveness of various components in OSAR.
We will first discuss the attention mechanism inside the AARN model, and second we will show the impact of the number of ROIs required for IDSN training.

\begin{figure}[h]
\begin{center} 
\subfigure[OSAR]{\label{Fig.abl_CT_ours2}
\includegraphics[width=0.45\linewidth]{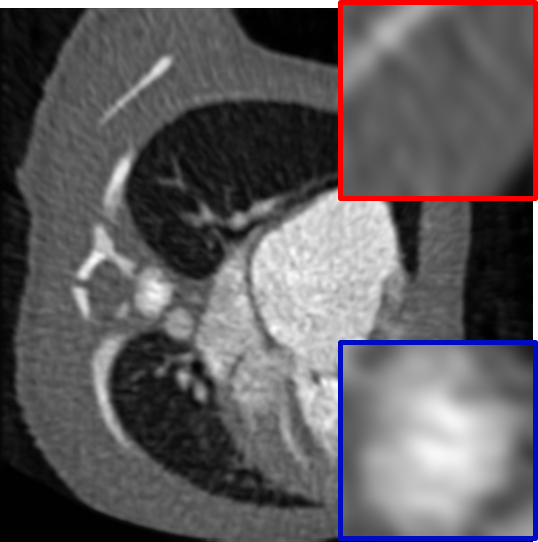}}
\subfigure[OSAR w/o attention]{\label{Fig.no_att2}
\includegraphics[width=0.45\linewidth]{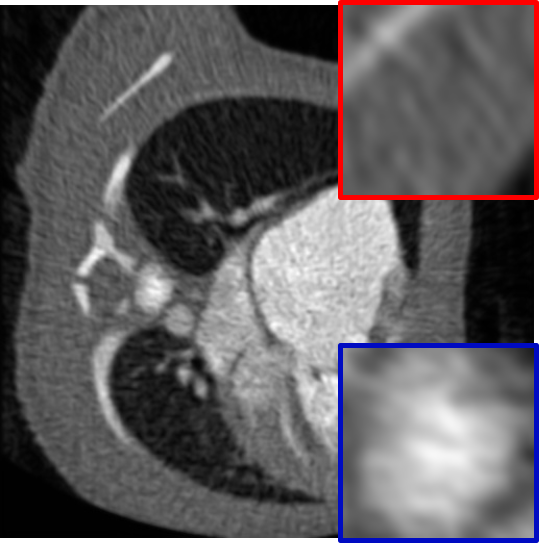}}
\caption{Results of OSAR, and OSAR (w/o attention) for the CT image.}
\label{Fig.wo}
\end{center}
\end{figure}

Fig. \ref{Fig.wo} and Table \ref{Table:wo} shows the qualitative and quantitative results when attention is removed. 
From the figure, we can see that the artifact in red region without attention is much more obvious compared with the proposed method.
Moreover, in the table, although the SNR improvement is larger when the attention mechanism is removed, it results in larger mean deviation (over 31.1\%) which the result will not be acceptable. This is because without attention, the model simply globally enhance the contrast of the image but did not preserve the tissue information. Thus, attention mechanism is essential to focus on specific regions. This justifies the additional complexity that the attention network brings. 

\begin{table}[h]
\setlength{\tabcolsep}{3pt}

\caption{Quantitative comparison of ablation study. Mean and Signal-to-Noise Ratio (SNR) for the largest homogeneous areas inside the marked regions of the CT images are reported.}
\label{Table:wo}
\centering
\begin{tabular}{cccccc}
\hline
Case & & Input & OSAR & OSAR (w/o attention)\\
\hline
\multirow{2}{*}{Fig.~\ref{Fig.wo} Red} & Mean & 58.8 & 59.7 & 77.1 \\ 
& SNR & 0.56 & 0.94 & 1.10\\
\hline
\multirow{2}{*}{Fig.~\ref{Fig.wo} Blue} & Mean & 71.8 & 63.5 & 88.8\\ 
& SNR & 0.53 & 1.02 & 1.14\\
\hline
\hline
\multirow{2}{*}{Overall} & $\Delta$Mean & 0.0\% & 11.6\% & 31.1\% \\ 
 & $\Delta$SNR & 0.0\% & 92.4\% & 115.0\%\\
\hline
\end{tabular}
\end{table}

\begin{figure}[h]
\begin{center}
\subfigure[CT image]{
\includegraphics[width=0.3\linewidth]{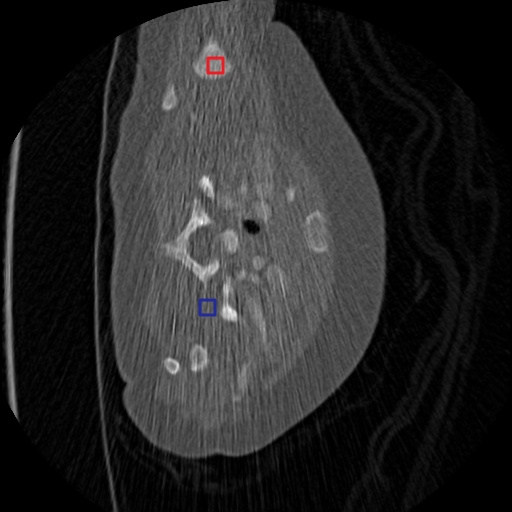}}
\subfigure[SNR/mean v.s. number of ROIs in (a)]{
\includegraphics[width=0.6\linewidth]{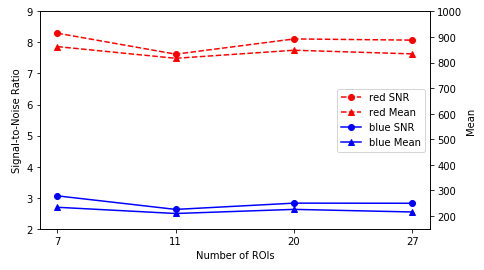}}
\subfigure[MRI image]{
\includegraphics[width=0.3\linewidth]{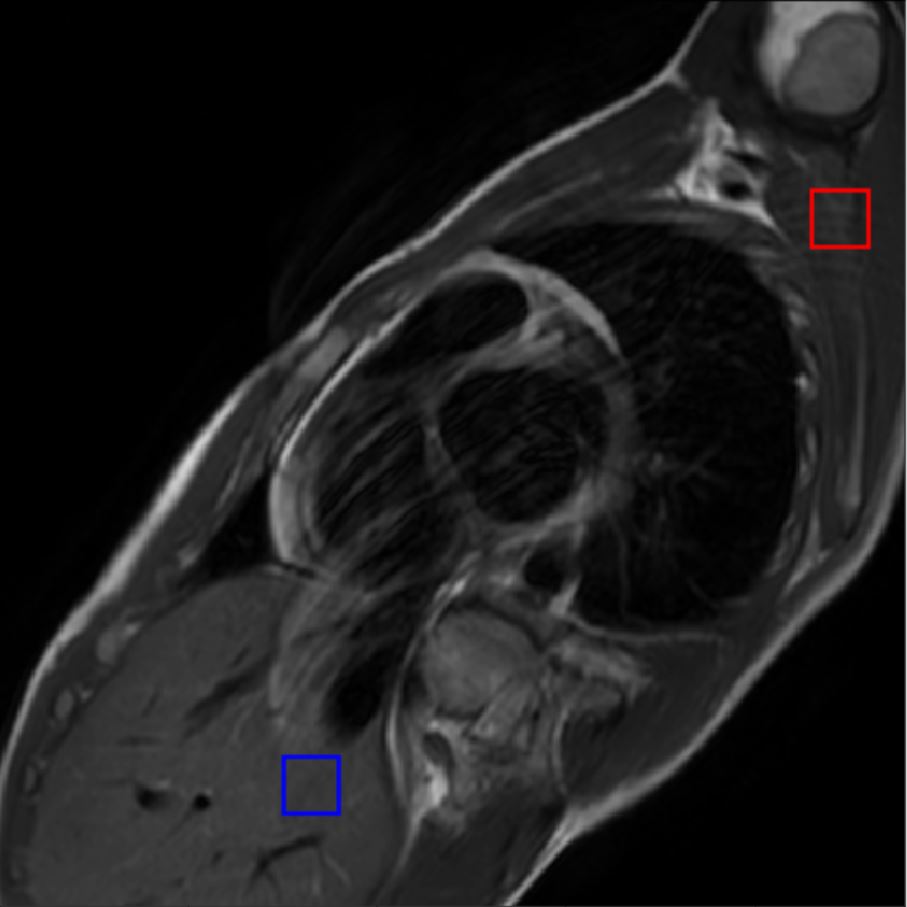}}
\subfigure[SNR/mean v.s. number of ROIs in (c)]{
\includegraphics[width=0.6\linewidth]{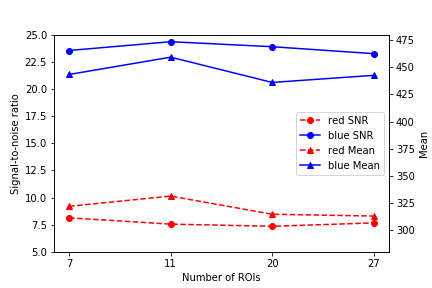}}

\caption{Impact of the number of annotated ROIs on artifact reduction quality of CT and MRI images.}
\label{Fig.roi}
\end{center}
\end{figure}

To examine the impact of the number of ROIs on the final artifact reduction quality, we apply our method to both CT (Fig. \ref{Fig.roi}(a)) and MRI (Fig. \ref{Fig.roi}(c)) images. In Fig. \ref{Fig.roi}(b) and (d), we collect the Signal-to-Noise Ratio (SNR) and mean value (substance information) in both red and blue regions marked in Fig. \ref{Fig.roi}(a) and (c), respectively. Details about these metrics can be found in Section \ref{sec:experiments}. We can observe that increasing the number of annotated ROIs from 7 to 27 results in almost the same artifact reduction quality for both CT and MRI. As such, only a small number of annotated ROIs are needed to achieve sufficiently good results. Moreover, the annotation usually takes less than a minute and is much faster compared with the artifact reduction time.

\section{Conclusions}
\label{sec:conclusions}
In this paper, we introduced OSAR, a ``One-Shot'' medical image artifact reduction framework, which exploits the power of deep learning to suppress additive artifacts in an image without using pre-trained networks. Unlike previous state-of-the-art methods which only learned the artifacts contained inside the training data, our method can be adapted for almost any medical images that contain varying additive artifacts.
Moreover, in order to fit clinical use, our network requires shorter runtime to obtain the denoised results than state-of-the-art.
Experimental results on cardiac CT and MRI images have shown that our framework can reduce additive noises and motion artifacts both qualitatively and quantitatively better than state-of-the-art.

\end{document}